\documentclass[a4paper,11pt]{article}
\usepackage{aaskaiid}
\usepackage{orcidlink}
\usepackage{xspace}
\usepackage[dvipsnames]{xcolor}
\usepackage{upgreek}
\usepackage{sidecap}

\def\gb{$\upgamma$B\xspace}
\def\gbs{$\upgamma$Bs\xspace}

\def\uJy{\upmu\text{Jy~beam}\xspace}
\def\uJybeam{\upmu\text{Jy~beam}^{-1}\xspace}

\title{Enabling population studies on wind-driven Galactic binary systems}
\ShortTitle{Wind-Driven Binary Systems}

\author[1,2]{B.~Marcote\orcidlink{0000-0001-9814-2354}}
\ShortName{Marcote et al.} 
\author[3]{P.~Benaglia\orcidlink{0000-0002-6683-3721}}
\author[4]{V.~Bosch-Ramon\orcidlink{0000-0002-6043-5079}}
\author[5]{M.~De~Becker\orcidlink{0000-0002-1303-6534}}
\author[6]{S.~del~Palacio\orcidlink{0000-0002-5761-2417}}
\author[2]{P.~Atri\orcidlink{0000-0001-8125-5619}}
\author[7]{J.~Mold\'on\orcidlink{0000-0002-8079-7608}}
\author[4]{J.~M.~Paredes\orcidlink{0000-0002-1566-9044}}
\author[4]{M.~Rib\'o\orcidlink{0000-0002-9931-4557}}
\author[3,8]{G.~E.~Romero\orcidlink{0000-0002-5260-1807}}
\author[9]{A.~Tej}

\affiliation[1]{Joint Institute for VLBI ERIC, Oude Hoogeveensedijk 4, 7991~PD Dwingeloo, The Netherlands}
\emailAdd{marcote@jive.eu}
\affiliation[2]{ASTRON, Netherlands Institute for Radio Astronomy, Oude Hoogeveensedijk 4, 7991~PD Dwingeloo, The Netherlands}
\affiliation[3]{Instituto Argentino de Radioastronomía (CONICET–CICPBA–UNLP), C.C. Nro 5, 1894, Villa Elisa, Argentina}
\affiliation[4]{Departament de Física Quàntica i Astrofísica, Institut de Ciències del Cosmos (ICCUB), Universitat de Barcelona, IEEC-UB, Martí i Franquès 1, 08028, Barcelona, Spain}
\affiliation[5]{Space Sciences, Technologies and Astrophysics Research (STAR) Institute, University of Liège, Quartier Agora, 19c, Allée du 6 Août, B5c, 4000, Sart Tilman, Belgium}
\affiliation[6]{Department of Space, Earth and Environment, Chalmers University of Technology, 412 96, Gothenburg, Sweden}
\affiliation[7]{Instituto de Astrofísica de Andalucía (CSIC), Glorieta de la Astronomía s/n, 18008, Granada, Spain}
\affiliation[8]{Facultad de Ciencias Astronómicas y Geofísicas, Universidad Nacional de La Plata, B1900FWA, La Plata, Buenos Aires, Argentina}
\affiliation[9]{Indian Institute of Space Science and Technology, Thiruvananthapuram, 695 547, Kerala, India}

\abstract{Galactic binaries driven by stellar wind shocks, such as colliding wind binaries (CWBs) and gamma-ray binaries (\gbs), harbor one of the most efficient particle acceleration engines known in the Universe. Despite their potential, these sources remain relatively unexplored, particularly in the domains of low radio frequencies and very high resolution. As a result, we lack comprehensive population studies and well-characterized individual systems. Only a few of these binaries, such as the iconic \gb PSR~B1259$-$63 or the massive CWB WR~140, have been studied in enough detail to probe their wind dynamics and shock physics. Current observations lack the sensitivity to detect weak non-thermal synchrotron emission from low-energy particle populations and the angular resolution to resolve shock structures on sub-au scales. The Square Kilometre Array Observatory (SKAO) will mark a significant improvement in both sensitivity and resolution with its SKA-low and SKA-mid telescopes, solving these challenges. This will enable systematic studies of the winds and shock interactions in these binary systems. Additionally, SKA-VLBI will facilitate the observation of changes in shock geometry at different orbital phases, linking particle acceleration processes to the binary's orbital characteristics and stellar wind properties.
SKAO will pave the way for comprehensive population studies of these energetic binary systems.}

\begin{document}

\maketitle

\section{Science background}

Stellar binary systems are widespread throughout our Galaxy. Among these, those that host a massive star are particularly significant, as they have the potential to be associated with high-energy phenomena \citep{Dubus2006,Dubus2013,DeBecker2013,DeBecker2017}.
These Galactic high-energy binaries represent unique laboratories for studying the mechanisms of particle acceleration in astrophysical environments. This chapter focuses on those binaries hosting massive stellar companions that are capable of efficiently accelerating particles to relativistic energies via shocks. Such accelerated particles are responsible for non-thermal emission across the entire electromagnetic spectrum (from radio to very high-energy $\upgamma$ rays, $\gtrsim 0.1~\text{TeV}$, in some cases), and their radio signatures provide crucial insights into the underlying physical processes.

A prominent class within this category of non-thermal emitters is that of colliding wind binaries (CWBs; \citealt{DeBecker2013}), in which the supersonic winds of two massive stars collide, generating strong shocks and non-thermal emission. The resulting interactions produce synchrotron radiation that is observable in the radio regime, offering a window into the stellar winds, magnetic fields, and relativistic electron populations in these systems. Closely related in their observational consequences, albeit with different evolutionary pathways and companion types, are (high-mass) gamma-ray binaries (\gbs; \citealt{Dubus2013}). These systems consist of a massive star and a compact companion, assumed to be a young non-accreting neutron star in all cases, where the stellar wind and the relativistic neutron star wind interact to produce very-high-energy radiation. In both classes, the study of the radio emission allows us to constrain the interplay between particle acceleration and the dynamics of the colliding flows. We note that other types of high-energy binaries that are not wind-powered dominated, such as X-ray binaries or novae, are covered in other Chapters (\citealt{TaoAn01.2026.SKA,Beri01.2026.SKA,Lico01.2026.SKA}).

This chapter explores the physics of these high-energy binary systems and the advancements that the Square Kilometre Array Observatory (SKAO) will produce in the field. By focusing on collisions between stellar winds, we examine the mechanisms driving the most efficient particle acceleration engines. We begin with an overview of colliding wind binaries and $\upgamma$-ray binaries, providing a coherent framework for understanding how these extreme environments contribute to the Galactic population of particle accelerators. We then discuss the new observational window opened by the SKAO for these fields.

\subsection{Colliding wind binaries: laboratories of wind shock physics}

\begin{SCfigure}[][t]
\centering
\includegraphics[width=0.45\columnwidth]{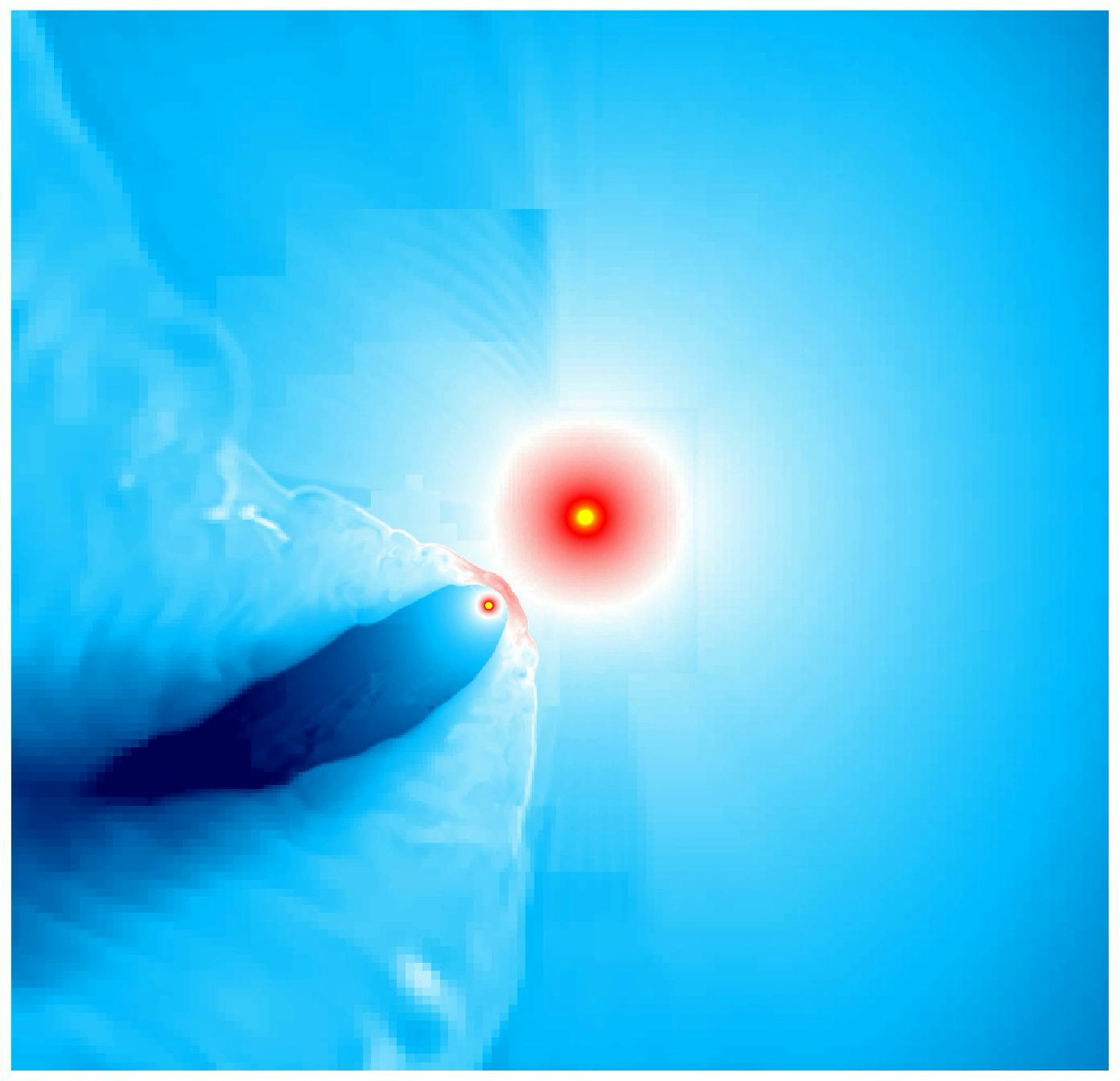}
\caption{Simulation of the interaction between the two stellar winds in a PACWB. Note the wind collision region formed between the two stars. Credit: Univ.\ of Liège/E.~R.~Parkin and E.~Gosset.}
\label{fig:cwbs}
\end{SCfigure}

CWBs \citep{DeBecker2013} consist of two massive stars (typically O or B spectral type, Wolf-Rayet, or Luminous Blue Variable stars) whose winds interact. The subset of these systems displaying evidence for efficient particle acceleration are known as Particle Accelerating Colliding Wind Binaries (PACWBs; \citealt{DeBecker2017}). In these systems, the two stellar winds collide, creating a shock interface where particle acceleration occurs via diffusive shock acceleration (DSA; see Figure~\ref{fig:cwbs}), and producing non-thermal emission that can be detectable from radio to $\upgamma$ rays.
The massive stars in these systems feature winds with velocities of $1\,000$--$3\,000\,\text{km\,s}^{-1}$ and mass loss rates of $\sim 10^{-7}\text{--}10^{-4}\,\text{M}_{\odot}\,\text{yr}^{-1}$, providing a large kinetic power and suitable conditions for efficient particle acceleration \citep[see e.g.][]{Pittard2021,delPalacio2023}.

Current observations reveal that these systems produce detectable synchrotron emission at radio wavelengths, which varies with orbital phase as the geometry of the wind collision region changes \citep{Dougherty2005}. X-ray observations have provided complementary information about the thermal properties of the shocked gas and, in some cases, evidence for non-thermal X-ray emission \citep{Hamaguchi2018, delPalacio2023}. The efficiency of particle acceleration in these systems is critically dependent on the wind momentum rate ratio ($\eta$), which determines the geometry of the wind collision region \citep{Eichler1993}.

The emission from these regions extends to scales of tens of au, which implies angular scales on the order of tens of milliarcseconds given the typical distances of a few kpc. This implies that only very long baseline interferometry (VLBI) observations can spatially resolve such regions. Direct imaging of these structures allows us to directly study the shocks where particle acceleration occurs and infer the properties of the two stellar winds. However, with current sensitivity, only the brightest, nearest PACWBs can be targeted with adequate signal-to-noise \citep{Dougherty2005,Benaglia2015,Marcote2021}. For fainter systems, like e.g.\ HD~167971 and HD~168112 in \citet{DeBecker2024}, current VLBI images do not provide enough sensitivity and image fidelity to constrain the morphology of the region and, consequently, the wind properties.

The broad population of CWBs thus remains poorly constrained, with estimates suggesting dozens to hundreds of such systems in our Galaxy, but only a few tens displaying confirmed non-thermal radio emission and an even smaller subset imaged with VLBI \citep{DeBecker2017}. In practice, this means that our current view of PACWBs is heavily biased toward the most luminous tail of the population, and the conditions for particle acceleration in more typical, lower-luminosity systems are essentially unknown.

\subsection{Gamma-ray binaries: the most extreme Galactic accelerators}

\begin{figure}[t]
\centering
\includegraphics[width=0.48\textwidth]{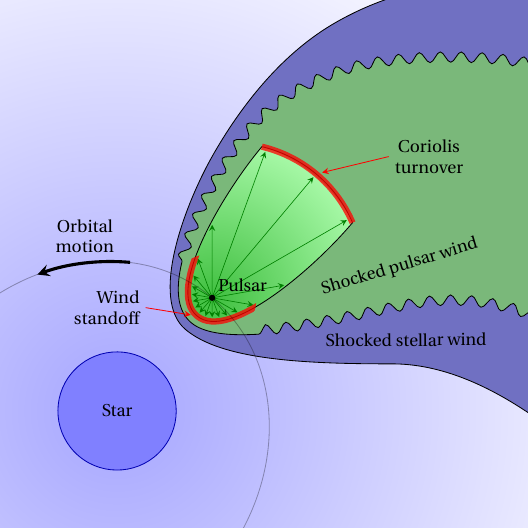}
\includegraphics[width=0.48\textwidth]{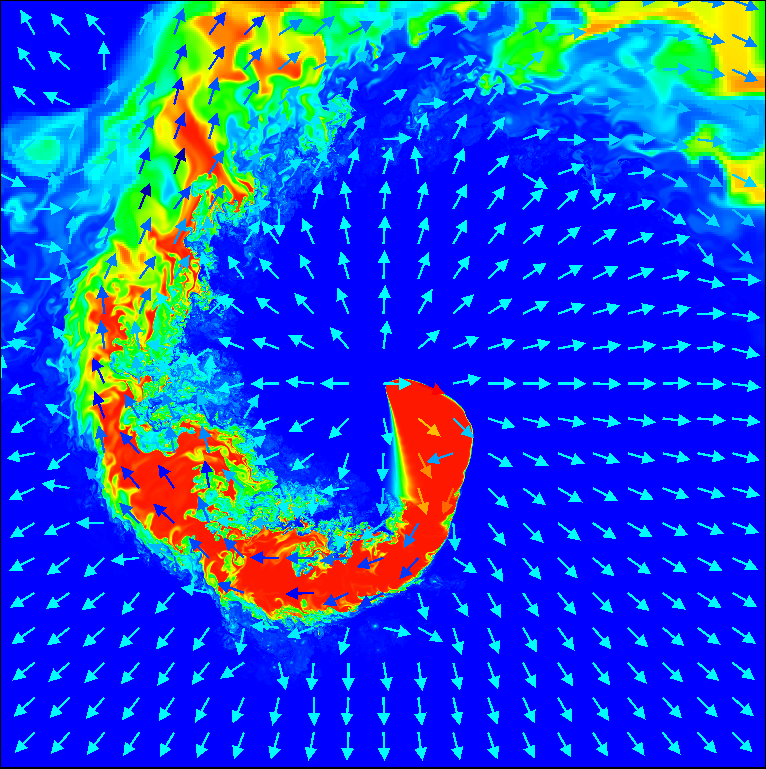}
\caption{{\em Left:} Artistic representation of the interaction between the stellar wind and the relativistic neutron star wind in a \gb \citep{Zabalza2013}. {\em Right:} same scenario as predicted by simulations, showing the larger scale interaction \citep{BoschRamon2012}.}
\label{fig:gbs}
\end{figure}

Gamma-ray binaries (\gbs; \citealt{Dubus2013}) represent an evolved and more extreme class of wind-interacting systems, where one of the massive stars has undergone a supernova explosion, leaving behind a compact object that continues to interact with the wind of the remaining massive star. These systems show emission from radio to very-high-energy $\upgamma$ rays, with the non-thermal spectral energy distribution (SED) being dominated by $\upgamma$ rays above 1~MeV. The interaction here is thought to be produced through shocks between the stellar wind of the massive star and the relativistic wind of a putative young non-accreting neutron star in all cases (see Figure~\ref{fig:gbs}).

Only ten \gbs have been discovered so far, with a wide range of orbital configurations and with a neutron star confirmed as the compact object in only a few of them \citep{Chernyakova2020,Bordas2024}. The radio emission in \gbs is produced by non-thermal synchrotron radiation and is often highly variable and orbitally modulated, showing outburst-like emission in some binaries. For the systems studied in detail, extended radio structures suggestive of cometary tails shaped by the interaction of the two winds are resolved via VLBI observations \citep{Massi2001,Dhawan2006,Moldon2011,Moldon2012,MillerJones2018}. These high-resolution radio observations have revealed that these extended structures evolve over the orbit, tracing the dynamics of the wind-wind shock and providing clues to the efficiency of particle acceleration far from the binary orbital scales.
Current observations reveal the presence of an unresolved core of unclear origin on milliarcsecond (or sub-au) scales that is responsible for most ($\sim 90\%$) of the emission \citep[see e.g.][]{Moldon2012}, which has been suggested to arise from the region around the Coriolis turnover \citep{BoschRamon2012}, where the shocked pulsar wind material changes direction due to orbital motion effects. The extended structure is thus significantly dimmer as it only covers the remaining $\sim 10\%$ emission.

Both classes of binary systems provide unique insights into shock physics and particle acceleration mechanisms. However, the study of these systems has been limited to a handful of sources. The increase in sensitivity and resolution of SKAO and SKA-VLBI will allow us to conduct systematic surveys and temporally-resolved studies of much fainter systems, thereby transforming individual, case-by-case investigations into genuine population studies.

\section{A wider observational window opened by the SKAO}

The SKAO --- especially in its AA4 baseline configuration --- will deliver a radical boost in sensitivity for radio observations in the $\sim 0.1\text{--}15~\text{GHz}$ frequency range. In its SKA-VLBI guise, where the SKA is integrated as a highly sensitive element into existing VLBI networks such as the European VLBI Network (EVN) and/or the Australian Long Baseline Array (LBA), it will play a major role in studying these systems. This integration will allow the observation of a substantially larger number of sources with higher resolution and sensitivity. Additionally, standalone SKA-low and SKA-mid will also be relevant to this chapter thanks to the achieved sensitivity (allowing monitoring programs with minimal observing time) and large field of view (to unveil new systems via commensal searches).

SKAO will thus permit not only the detection of much fainter sources, but also the precise mapping of spatial structures and temporal evolution in wind collisions at sub-milliarcsecond scales. This is crucial for understanding the dynamics of binary interactions and particle acceleration regions. In the following, we discuss the enhanced capabilities of SKA-VLBI in resolving shock structures and explore how SKAO observations (using both SKA-low and SKA-mid) facilitate efficient monitoring and discovery of new wind-driven binaries.

\subsection{Resolving the shocks with SKA-VLBI}

\begin{figure}[t]
\centering
\includegraphics[width=0.45\columnwidth]{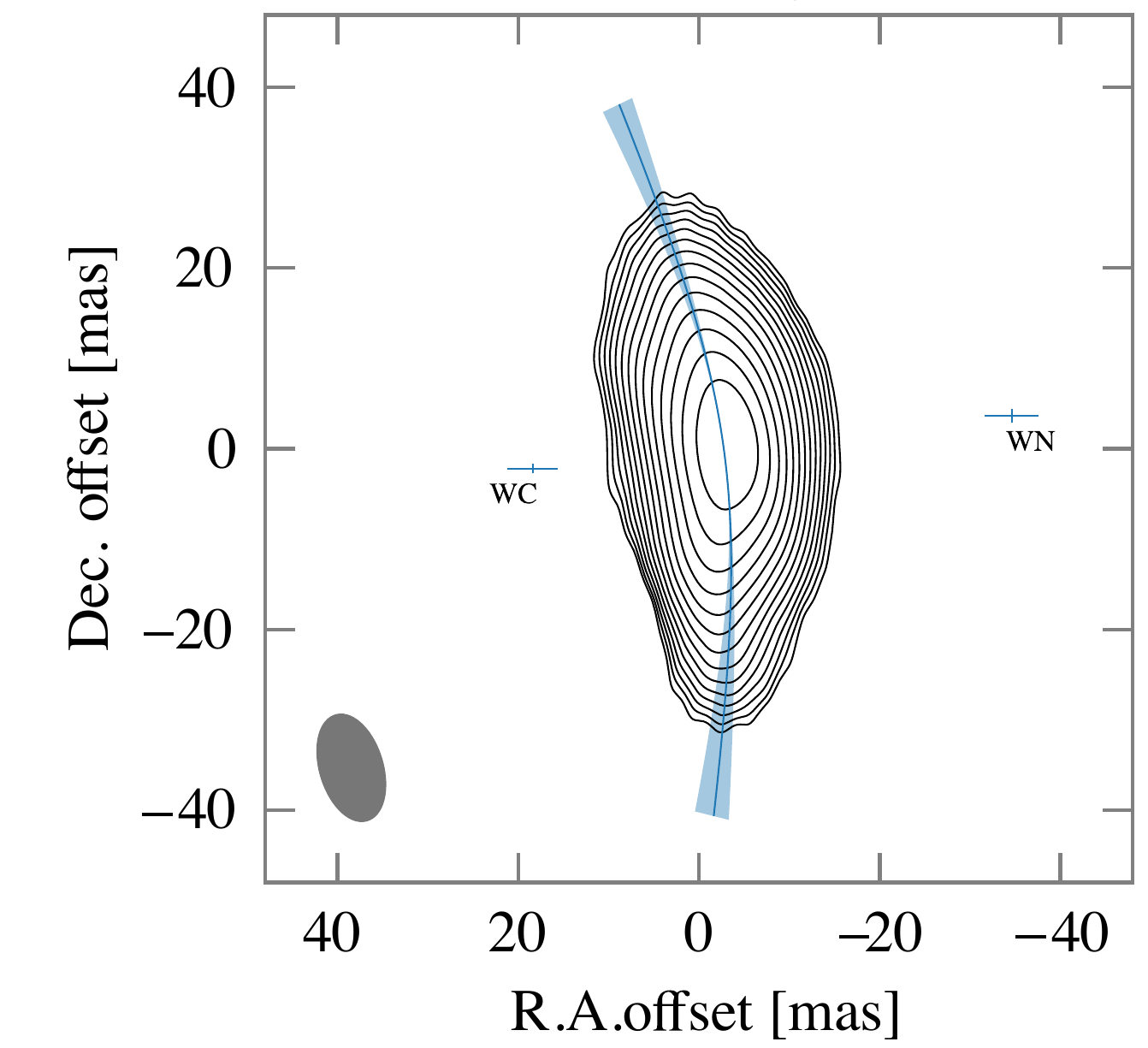}
\includegraphics[width=0.45\columnwidth]{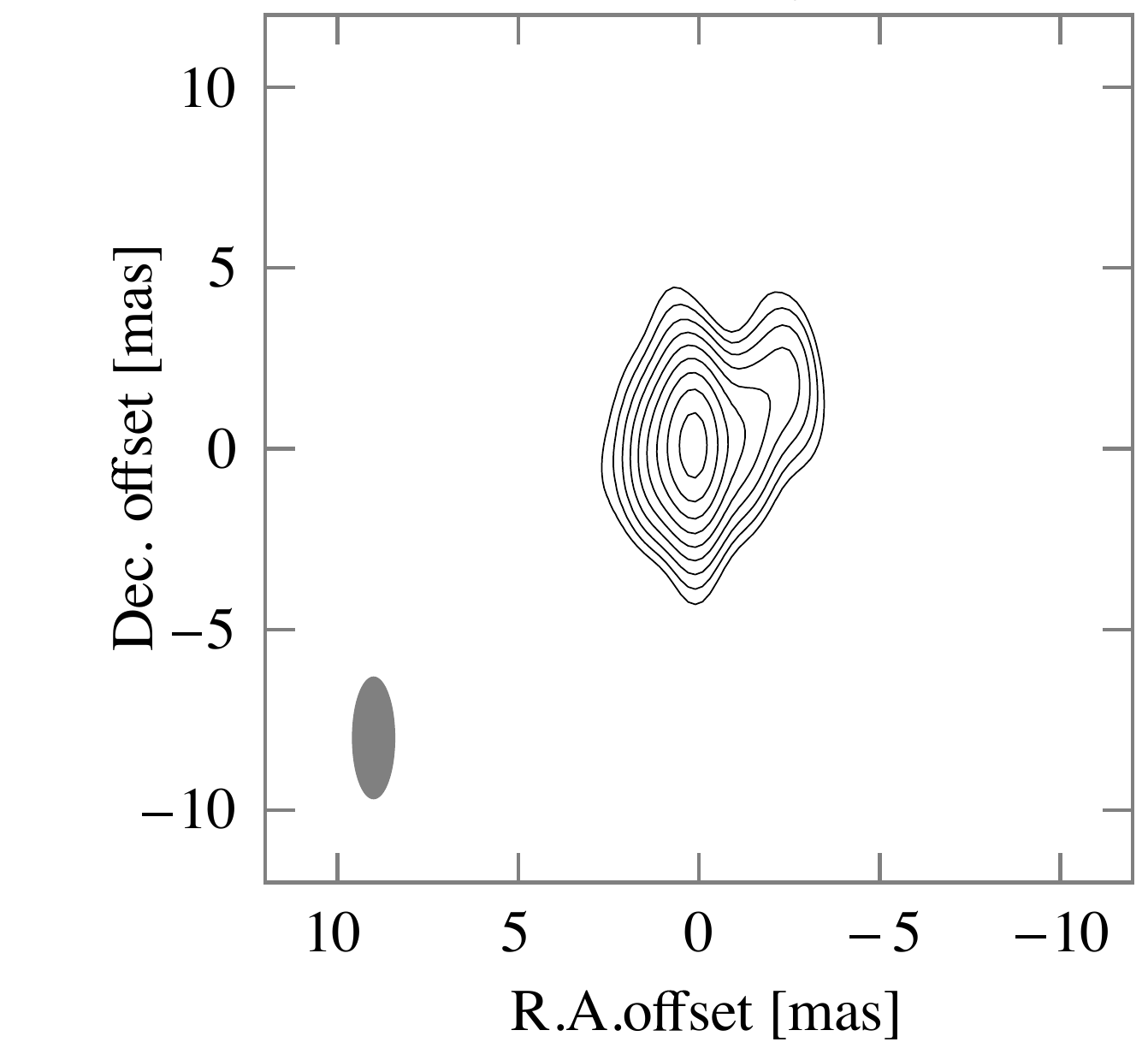}
\caption{{\em Left:} Radio emission from the PACWB Apep \citep[adapted from][]{Marcote2021}. Note the position of the two stars in the system (WC and WN). {\em Right}: Radio emission from the \gb LS~5039 \citep{Moldon2012}. In this case the entire system is well within the core emission (orbital size of $< 0.1~\text{mas}$).}
\label{fig:vlbi-images}
\end{figure}

SKA-VLBI represents a significant step forward in our ability to study high-energy Galactic binaries with unprecedented sensitivity and resolution. In the AA4 configuration, SKA-mid dishes phased up for VLBI will become the most sensitive (equivalent) dish in any existing VLBI network, yielding image rms levels of a few $\uJybeam$ in $\lesssim 1$~h at GHz frequencies for typical VLBI configurations.
Even without increasing baseline lengths compared to current VLBI networks, SKA-VLBI's improvement in sensitivity will dramatically enhance our ability to trace the most extended emission components and resolve the innermost parts of the emission (as done for the brighter systems to date, see Figure~\ref{fig:vlbi-images}), as demonstrated by the first MeerKAT-EVN tests\footnote{\url{https://jive.eu/news/earth-sized-radio-observatory-just-got-better-south-africas-meerkat-telescope-joins-forces}.}.

Current VLBI observations of \textbf{PACWBs} have been limited to only the brightest and most extreme systems due to sensitivity constraints. The handful of sources that have been characterized with VLBI techniques, such as WR~140 \citep{Dougherty2005}, HD~93129A \citep{Benaglia2015}, or Apep \citep{Marcote2021}, represent exceptional cases where the synchrotron emission from wind collision regions is sufficiently bright to be properly mapped with existing facilities. These observations have provided crucial insights into the morphology and orbital variability of wind-wind collision regions, and the stellar wind characteristics, but they sample only the most luminous end of the PACWB population.

SKA-VLBI will fundamentally change this limitation. The expected improvement in the sensitivity of continuum data of about one order of magnitude on long baselines (those including the phased-up SKA core) is particularly relevant for mapping the fainter parts of the emission and hence the curvature of the regions. Higher image fidelity will translate into a more accurate reconstruction of the bow-shaped structure of the emitting region, which directly leads to tighter constraints on the wind-momentum rate ratio and hence on the individual mass-loss rates and wind velocities. For systems with present-day VLBI detections at the mJy level and fractional uncertainties of $\gtrsim 20\%$ on $\eta$ \citep[e.g.][]{Dougherty2005}, such a boost in sensitivity would improve the effective resolution (as in the previous MeerKAT-EVN example) and likely reduce uncertainties on $\eta$ to the $\lesssim 1\%$ level, enabling meaningful comparisons with detailed hydrodynamical and stellar models across a significantly larger sample of systems.

Most importantly, SKA-VLBI will unveil the PACWB nature of the bulk of the population by making it possible to resolve wind collision shocks in sources with flux densities of only tens of $\uJybeam$. Assuming image rms levels of a few $\uJybeam$ in observations lasting a few hours, PACWBs with peak brightnesses of $30$--$50~\uJybeam$ will become accessible to robust imaging, whereas such sources are currently below the practical detection threshold of most VLBI arrays. This capability will finally allow detection and characterization of the low-luminosity sources that likely dominate the PACWB population but have remained inaccessible to current VLBI networks, yielding the first statistically robust population studies of wind-collision geometries and particle acceleration efficiencies. Furthermore, we would also enter the regime where thermal emission from the individual stars would be detectable ($\sim 10~\uJybeam$), and hence VLBI maps will be able to detect at once the two stellar components plus the shock (directly providing a 2D orbital model of the system).

In the field of \gbs, the observed extended radio emission exhibits characteristic cometary tail-like structures that trace the evolution of the shocked material along the orbit as it propagates away from the system. Previous VLBI observations have revealed these structures in a handful of systems, such as PSR~B1259$-$63 \citep{Moldon2011}, LS~I~+61~303 \citep{Massi2001,Dhawan2006}, LS~5039 \citep{Moldon2012}, or HESS~J0632+057 \citep{Moldon2011hess,Marcote2014}. While the core emission is typically strong (on the order of a-few-to-tens of mJy), tracing the most extended parts and their morphologies is limited due to the fidelity obtained in current images and/or by the observing time that such campaigns require (tens of epochs of $\sim 10~\text{h}$ each).

This enhanced sensitivity will enable detailed mapping of the evolution of shocked material as it flows away from the binary system, following previous works but over a larger range of spatial scales and orbital phases. The improved detection capability will allow systematic monitoring of the morphological changes in the extended emission throughout complete orbital cycles, providing crucial constraints on the physical conditions in the shock regions and the transport of relativistic particles. Furthermore, SKA-VLBI observations at higher frequencies ($\sim 15~\text{GHz}$) will help to resolve the long-standing question regarding the origin of the bulk of the radio emission in \gbs. The higher angular resolution and sensitivity of SKA-VLBI at these frequencies will allow us to probe smaller angular scales at comparable or better signal-to-noise, and to search for subtle structural changes expected near the Coriolis turnover region along the orbit. In particular, mapping the frequency-dependent size and position of the core component over multiple epochs will provide stringent tests of models in which the bulk of the radio emission originates in this region rather than in the immediate vicinity of the compact object.

Sensitive SKA-VLBI observations at the lowest frequencies ($\sim 100~\text{MHz}$, with SKA-Low antennas) would also allow us to explore, for the first time, the existence of extended (halo-like) emission on larger scales. It is expected \citep[see e.g.][]{BoschRamon2011} that the spiral-shaped structure formed by the shocked winds becomes disrupted after several orbital cycles at distances of $\gtrsim 1\,000~\text{au}$ (or arcsecond-to-arcmin scales given the typical distances to the known \gbs). At such distances, there should still be enough energy to accelerate particles that could produce emission at these frequencies. Detecting and characterizing such halos would provide a direct link between the small-scale wind interaction region accessible to VLBI and the larger-scale injection of relativistic particles into the surrounding interstellar medium.

\subsection{Monitoring of binaries with SKAO}

Systematic monitoring of both PACWBs and \gbs with snapshot observations using SKAO will provide unprecedented opportunities to confirm the nature of these sources through detailed light-curve analyses and spectral evolution studies. The enhanced sensitivity of SKAO will enable routine detection of orbital modulation and spectral evolution in the radio emission of these binary systems, using short observations that can be efficiently scheduled over many orbital cycles. The sensitivity of SKAO enables this kind of monitoring for tens of sources with minimal observing time (tens of minutes to a few hours). As demonstrated in recent studies of systems like HD~93129A \citep{Benaglia2025}, intensive radio monitoring during key orbital phases allows tracking of the physical conditions in wind-collision regions, revealing how synchrotron emission varies with orbital separation and wind density.

The critical advantage of SKAO monitoring lies in its ability to characterize the spectral turnover frequency, which typically occurs at frequencies of $0.5\text{--}1~\text{GHz}$ for the studied systems, and its evolution across the orbital cycle \citep{Marcote2015ls5039,Marcote2016lsi,Bloot2022,delPalacio2022,Tasseroul2025}. This turnover frequency is governed by a combination of synchrotron self-absorption, the Razin effect (dominant in \gbs), and free-free absorption (for PACWBs), with the exact mechanism varying between sources and orbital configurations. By modeling these spectral turnovers and their temporal evolution, SKAO observations will provide direct constraints on the magnetic field strength, electron density, and mass-loss rates in the wind-collision regions. Furthermore, the precise timing and spectral evolution of these features will allow the characterization of orbital motion and wind structure, distinguishing between different theoretical models for particle acceleration in these extreme environments.

\subsection{Discovery of binaries through commensal searches}

Searches for variable radio sources in SKAO archival observations have the potential to uncover a previously undetected population of PACWBs within our Galaxy. The current population of detected PACWBs is skewed towards regions with minimal (optical) extinction and relatively short orbital periods, as optical catalogs struggle to identify the orbital motions of systems with periods $\gg 100~\text{yr}$. However, radio observations uniquely favor the detection of PACWBs with long orbital periods of tens-to-hundreds of years. In such cases, the large binary separation leads to wind interaction regions sufficiently far from the stars, making the non-thermal synchrotron radio emission much more likely to escape unabsorbed due to free-free absorption generated by the stellar winds. These wide binaries remain elusive to optical and high-energy data, as their large separations produce minimal radial velocity shifts and weaker or no high-energy emission. SKAO’s sensitive radio surveys will thus selectively reveal this population, allowing, for the first time, a complete census of PACWBs across the full range of orbital parameters.

The defining radio characteristics that will enable identification of PACWBs in such commensal searches include: (i) non-thermal radio spectra with negative spectral indices (typically $\alpha \sim -0.5$ to $-1$, where $S_\nu \propto \nu^\alpha$) indicative of synchrotron emission; (ii) moderate flux variability on timescales from months to decades, reflecting changes in the wind collision region over the orbit; and (iii) spatial association with massive stars or stellar clusters that host evolved, wind-driving objects. By combining SKAO continuum surveys with optical/infrared catalogs of massive stars and {\em Gaia} astrometry, candidate PACWBs can be efficiently selected for targeted follow-up, even when their orbital periods far exceed the duration of current spectroscopic monitoring programs.


\section{Scientific Implications}

\subsection{Unveiling a more complete population of PACWBs}

As mentioned above, the improvement in sensitivity and survey capabilities of SKAO would allow us to unveil a previously hidden population of PACWBs, fundamentally broadening our understanding of massive star interactions in the Galaxy. A substantial fraction of the wide, long-period systems that remain undetected should be discovered within the first decade of SKAO operations. Steady or slowly variable non-thermal radio emission at the tens of $\uJy$ level should be detectable in deep all-sky surveys, even in regions of high optical extinction where traditional spectroscopic searches are incomplete.

SKAO’s sub-$\upmu$Jy continuum sensitivity and broadband coverage will also make it possible to disentangle the thermal and non-thermal components, detecting the weak non-thermal population tail. All this will provide a complete census of Galactic PACWBs, including systems that radiate below current instrumental thresholds, establishing the limiting conditions necessary for particle acceleration in these wind collisions. In combination with VLBI follow-up for a representative subsample, this will allow us to connect global properties such as orbital period and stellar type with local shock conditions and particle acceleration efficiency across the full PACWB parameter space.


\subsection{PACWB implications in the Galactic cosmic-ray background}

Unveiling the extended population of PACWBs in our Galaxy will enable us to directly constrain their role in the diffuse Galactic cosmic-ray budget. These observations will provide empirical estimates of the acceleration and re-acceleration efficiencies of relativistic particles in these binaries. When combined with hydrodynamical models, these data will define the contribution of stellar-wind termination shocks to the low-energy cosmic-ray population. Population synthesis will refine the total energy injection rate of PACWBs into the interstellar medium (ISM), bridging the gap between isolated stellar wind shocks and supernova remnants as distributed sources of Galactic cosmic rays. Current estimates suggest that PACWBs can contribute up to a few percent of this background, making them a testable source of cosmic rays \citep{DeBecker2017,Seo2018,Kalyashova2019,Wang2022}.

Population studies will also provide insights into the evolutionary connection between PACWBs and \gbs. The relative numbers and characteristics of these populations will constrain evolutionary models and rates, providing a more complete picture of the life cycle of massive binary systems. In particular, establishing whether most \gbs descend from PACWBs with specific ranges of orbital separations and wind parameters will shed light on how supernova kicks and mass transfer episodes shape the subsequent binary configurations.

\subsection{Prospects on the origin of {\boldmath{$\upgamma$}}Bs and their emission}

SKA-VLBI capabilities will resolve the extended synchrotron structures formed in \gbs over complete orbital cycles. High-dynamic-range imaging will map the spiral and cometary tail morphologies predicted from magnetohydrodynamic simulations \citep{BoschRamon2012}. The combination of temporal-evolution radio morphology and spectral evolution will constrain key parameters such as mixing between the two winds, particle advection speed, and cooling timescales. These observations will serve as benchmarks for 3D relativistic simulations of binary shocks, providing the first empirical tests beyond the one-zone models.

The enhanced sensitivity and resolution would allow us to constrain, for the first time, the origin of the bulk of the radio emission, and thus put the first observational constraints on the Coriolis turnover region. By measuring how the size, position, and spectrum of the compact core component vary with orbital phase and frequency, SKA-VLBI will be able to distinguish between scenarios in which the main radio-emitting region is located close to the stagnation point of the colliding winds or further downstream near the Coriolis turnover, where the shocked flow is strongly bent by orbital motion.

Furthermore, these observations would also be fundamental to address the formation channels that led to the compact objects in \gbs.  SKA-VLBI observations would allow us to perform more precise measurements of the proper motions and parallaxes of these binaries \citep{Moldon2012lspsr,Marcote2018,Wu2018}. Constraining the kick velocities of these systems is fundamental to unveil the possible supernova remnants or clusters associated with these systems, and hence place better estimates on the lifetimes of \gbs \citep{CarreteroCastrillo2025}, similarly to what has been conducted for low-mass X-ray binaries \citep{atri2019}. This is particularly relevant for sources with high optical absorption that were not detected by the {\em Gaia} satellite.

At low frequencies, SKA(-VLBI)-low would have the potential to detect diffuse halos around \gbs created by the acceleration of low-energy electrons once the spiral-shaped structure formed by the shocked winds gets disrupted. Mapping their morphology and spectral steepening will quantify diffusion lengths, escape times, and the coupling between binary wind outflows and the local ISM. These measurements will allow us to better understand the cosmic-ray acceleration produced in these systems at large scales and their role as localized injectors of relativistic particles into the Galactic environment.

\subsection{Binaries in a multi-wavelength context}

The field of high-energy binary systems represents a unique opportunity to establish a multi\-wa\-ve\-length collaborative effort, as these sources exhibit emission across the entire electromagnetic spectrum. Simultaneous observations, particularly in the radio, X-ray, GeV, and TeV domains, are crucial for a comprehensive understanding of these systems, especially in the case of \gbs. On the other hand, radio observations of PACWBs would enable accurate predictions of these sources as potential high-energy emitters, which could be verified by current (e.g.\ with the MAGIC, H.E.S.S.\ or VERITAS telescopes) and upcoming facilities like the Cherenkov Telescope Array Observatory (CTAO). In this context, SKAO will provide the densely sampled, broadband radio light curves and spectra required to interpret high-energy variability and to disentangle leptonic and hadronic emission scenarios in detailed spectral modeling.

\section{Concluding remarks}

As we have seen, the SKAO will revolutionize the study of wind-driven high-energy binary systems, enabling systematic population studies and detailed observations of the most efficient particle acceleration mechanisms in our Galaxy. SKAO’s capabilities will provide crucial insights into the role of these systems in Galactic cosmic rays and advance our understanding of particle acceleration across the electromagnetic spectrum.

\bibliographystyle{abbrvnat-maxbibnames4}
\bibliography{chapter} 

@ARTICLE{Benaglia2015,
       author = {{Benaglia}, P. and {Marcote}, B. and {Mold{\'o}n}, J. and {Nelan}, E. and {De Becker}, M. and {Dougherty}, S.~M. and {Koribalski}, B.~S.},
        title = "{A radio map of the colliding winds in the very massive binary system HD 93129A}",
      journal = {\aap},
         year = 2015,
        month = jul,
       volume = {579},
          eid = {A99},
        pages = {A99},
          doi = {10.1051/0004-6361/201425595},
archivePrefix = {arXiv},
       eprint = {1503.07752},
 primaryClass = {astro-ph.SR},
       adsurl = {https://ui.adsabs.harvard.edu/abs/2015A&A...579A..99B}
}

@incollection{Lico01.2026.SKA, author = {Rocco Lico and author2 and author3 and author4 and author5},title = {},year = {2026},publisher = {},note = {arXiv search: Report number AASKAII/Lico01},booktitle = {Advancing Astrophysics with the SKA -- II (AASKAII)}}

@incollection{TaoAn01.2026.SKA, author = {Tao An and author2 and author3 and author4 and author5},title = {},year = {2026},publisher = {},note = {arXiv search: Report number AASKAII/TaoAn01},booktitle = {Advancing Astrophysics with the SKA -- II (AASKAII)}}

@incollection{Beri01.2026.SKA, author = {Aru Beri and author2 and author3 and author4 and author5},title = {},year = {2026},publisher = {},note = {arXiv search: Report number AASKAII/Beri01},booktitle = {Advancing Astrophysics with the SKA -- II (AASKAII)}}

@ARTICLE{atri2019,
       author = {{Atri}, P. and {Miller-Jones}, J.~C.~A. and {Bahramian}, A. and {Plotkin}, R.~M. and {Jonker}, P.~G. and {Nelemans}, G. and {Maccarone}, T.~J. and {Sivakoff}, G.~R. and {Deller}, A.~T. and {Chaty}, S. and et al.},
        title = "{Potential kick velocity distribution of black hole X-ray binaries and implications for natal kicks}",
      journal = {\mnras},
         year = 2019,
        month = nov,
       volume = {489},
       number = {3},
        pages = {3116-3134},
          doi = {10.1093/mnras/stz2335},
archivePrefix = {arXiv},
       eprint = {1908.07199},
 primaryClass = {astro-ph.HE},
       adsurl = {https://ui.adsabs.harvard.edu/abs/2019MNRAS.489.3116A}
}

@ARTICLE{DeBecker2024,
       author = {{De Becker}, M. and {Marcote}, B. and {Furst}, T. and {Benaglia}, P.},
        title = "{High-resolution radio imaging of the two particle-accelerating colliding-wind binaries HD 167971 and HD 168112}",
      journal = {\aap},
         year = 2024,
        month = feb,
       volume = {682},
          eid = {A160},
        pages = {A160},
          doi = {10.1051/0004-6361/202348622},
archivePrefix = {arXiv},
       eprint = {2401.02712},
 primaryClass = {astro-ph.SR},
       adsurl = {https://ui.adsabs.harvard.edu/abs/2024A&A...682A.160D}
}

@ARTICLE{Benaglia2025,
       author = {{Benaglia}, P. and {del Palacio}, S. and {Saponara}, J. and {Blanco}, A.~B. and {De Becker}, M. and {Marcote}, B.},
        title = "{Radio study of the colliding-wind binary HD 93129A near periastron and its surroundings}",
      journal = {\aap},
         year = 2025,
        month = jun,
       volume = {698},
          eid = {A23},
        pages = {A23},
          doi = {10.1051/0004-6361/202453422},
archivePrefix = {arXiv},
       eprint = {2503.11776},
 primaryClass = {astro-ph.SR},
       adsurl = {https://ui.adsabs.harvard.edu/abs/2025A&A...698A..23B}
}

@ARTICLE{Bloot2022,
       author = {{Bloot}, S. and {Callingham}, J.~R. and {Marcote}, B.},
        title = "{Radio modelling of the brightest and most luminous non-thermal colliding-wind binary Apep}",
      journal = {\mnras},
         year = 2022,
        month = jan,
       volume = {509},
       number = {1},
        pages = {475-488},
          doi = {10.1093/mnras/stab2976},
archivePrefix = {arXiv},
       eprint = {2110.06154},
 primaryClass = {astro-ph.SR},
       adsurl = {https://ui.adsabs.harvard.edu/abs/2022MNRAS.509..475B}
}

@ARTICLE{DeBecker2013,
       author = {{De Becker}, M. and {Raucq}, F.},
        title = "{Catalogue of particle-accelerating colliding-wind binaries}",
      journal = {\aap},
         year = 2013,
        month = oct,
       volume = {558},
          eid = {A28},
        pages = {A28},
          doi = {10.1051/0004-6361/201322074},
archivePrefix = {arXiv},
       eprint = {1308.3149},
 primaryClass = {astro-ph.HE},
       adsurl = {https://ui.adsabs.harvard.edu/abs/2013A&A...558A..28D}
}

@ARTICLE{Marcote2021,
       author = {{Marcote}, B. and {Callingham}, J.~R. and {De Becker}, M. and {Edwards}, P.~G. and {Han}, Y. and {Schulz}, R. and {Stevens}, J. and {Tuthill}, P.~G.},
        title = "{AU-scale radio imaging of the wind collision region in the brightest and most luminous non-thermal colliding wind binary Apep}",
      journal = {\mnras},
         year = 2021,
        month = feb,
       volume = {501},
       number = {2},
        pages = {2478-2486},
          doi = {10.1093/mnras/staa3863},
archivePrefix = {arXiv},
       eprint = {2012.06571},
 primaryClass = {astro-ph.SR},
       adsurl = {https://ui.adsabs.harvard.edu/abs/2021MNRAS.501.2478M}
}

@ARTICLE{Tasseroul2025,
       author = {{Tasseroul}, M. and {De Becker}, M. and {Blanco}, A.~B. and {Benaglia}, P. and {del Palacio}, S.},
        title = "{Foreground and internal free-free absorption in particle-accelerating colliding-wind binaries: Insights from the radio emission of WR147}",
      journal = {\aap},
         year = 2025,
        month = sep,
       volume = {701},
          eid = {A243},
        pages = {A243},
          doi = {10.1051/0004-6361/202555803},
archivePrefix = {arXiv},
       eprint = {2508.10506},
 primaryClass = {astro-ph.SR},
       adsurl = {https://ui.adsabs.harvard.edu/abs/2025A&A...701A.243T}
}

@ARTICLE{delPalacio2023,
       author = {{del Palacio}, S. and {Garc{\'\i}a}, F. and {De Becker}, M. and {Altamirano}, D. and {Bosch-Ramon}, V. and {Benaglia}, P. and {Marcote}, B. and {Romero}, G.~E.},
        title = "{Evidence for non-thermal X-ray emission from the double Wolf-Rayet colliding-wind binary Apep}",
      journal = {\aap},
         year = 2023,
        month = apr,
       volume = {672},
          eid = {A109},
        pages = {A109},
          doi = {10.1051/0004-6361/202245505},
archivePrefix = {arXiv},
       eprint = {2302.08170},
 primaryClass = {astro-ph.HE},
       adsurl = {https://ui.adsabs.harvard.edu/abs/2023A&A...672A.109D}
}

@ARTICLE{Hamaguchi2018,
       author = {{Hamaguchi}, Kenji and {Corcoran}, Michael F. and {Pittard}, Julian M. and {Sharma}, Neetika and {Takahashi}, Hiromitsu and {Russell}, Christopher M.~P. and {Grefenstette}, Brian W. and {Wik}, Daniel R. and {Gull}, Theodore R. and {Richardson}, Noel D. and {Madura}, Thomas I. and {Moffat}, Anthony F.~J.},
        title = "{Non-thermal X-rays from colliding wind shock acceleration in the massive binary Eta Carinae}",
      journal = {Nature Astronomy},
         year = 2018,
        month = jul,
       volume = {2},
        pages = {731-736},
          doi = {10.1038/s41550-018-0505-1},
archivePrefix = {arXiv},
       eprint = {1904.09219},
 primaryClass = {astro-ph.HE},
       adsurl = {https://ui.adsabs.harvard.edu/abs/2018NatAs...2..731H}
}

@ARTICLE{Eichler1993,
       author = {{Eichler}, D. and {Usov}, V.},
        title = "{Particle Acceleration and Nonthermal Radio Emission in Binaries of Early-Type Stars}",
      journal = {\apj},
         year = 1993,
        month = jan,
       volume = {402},
        pages = {271},
          doi = {10.1086/172130},
       adsurl = {https://ui.adsabs.harvard.edu/abs/1993ApJ...402..271E}
}

@ARTICLE{Dougherty2005,
       author = {{Dougherty}, S.~M. and {Beasley}, A.~J. and {Claussen}, M.~J. and {Zauderer}, B.~A. and {Bolingbroke}, N.~J.},
        title = "{High-Resolution Radio Observations of the Colliding-Wind Binary WR 140}",
      journal = {\apj},
         year = 2005,
        month = apr,
       volume = {623},
       number = {1},
        pages = {447-459},
          doi = {10.1086/428494},
archivePrefix = {arXiv},
       eprint = {astro-ph/0501391},
 primaryClass = {astro-ph},
       adsurl = {https://ui.adsabs.harvard.edu/abs/2005ApJ...623..447D}
}

@ARTICLE{delPalacio2022,
       author = {{del Palacio}, S. and {Benaglia}, P. and {De Becker}, M. and {Bosch-Ramon}, V. and {Romero}, G.~E.},
        title = "{The non-thermal emission from the colliding-wind binary Apep}",
      journal = {\pasa},
         year = 2022,
        month = jan,
       volume = {39},
          eid = {e004},
        pages = {e004},
          doi = {10.1017/pasa.2021.60},
archivePrefix = {arXiv},
       eprint = {2111.07442},
 primaryClass = {astro-ph.HE},
       adsurl = {https://ui.adsabs.harvard.edu/abs/2022PASA...39....4D}
}

@article{Dubus2013,
	adsurl = {http://adsabs.harvard.edu/abs/2013A%26ARv..21...64D},
	archiveprefix = {arXiv},
	author = {{Dubus}, G.},
	doi = {10.1007/s00159-013-0064-5},
	eid = {64},
	eprint = {1307.7083},
	journal = {\aapr},
	month = aug,
	pages = {64},
	primaryclass = {astro-ph.HE},
	title = {{Gamma-ray binaries and related systems}},
	volume = 21,
	year = 2013
}

@ARTICLE{Dubus2006,
       author = {{Dubus}, G.},
        title = "{Gamma-ray binaries: pulsars in disguise?}",
      journal = {\aap},
         year = 2006,
        month = sep,
       volume = {456},
       number = {3},
        pages = {801-817},
          doi = {10.1051/0004-6361:20054779},
archivePrefix = {arXiv},
       eprint = {astro-ph/0605287},
 primaryClass = {astro-ph},
       adsurl = {https://ui.adsabs.harvard.edu/abs/2006A&A...456..801D}
}

@article{DeBecker2017,
	adsurl = {https://ui.adsabs.harvard.edu/abs/2017A&A...600A..47D},
	archiveprefix = {arXiv},
	author = {{De Becker}, M. and {Benaglia}, P. and {Romero}, G.~E. and {Peri}, C.~S.},
	doi = {10.1051/0004-6361/201629110},
	eid = {A47},
	eprint = {1703.02385},
	journal = {\aap},
	month = apr,
	pages = {A47},
	primaryclass = {astro-ph.HE},
	title = {{An investigation into the fraction of particle accelerators among colliding-wind binaries. Towards an extension of the catalogue}},
	volume = {600},
	year = 2017}

@ARTICLE{Pittard2021,
       author = {{Pittard}, J.~M. and {Romero}, G.~E. and {Vila}, G.~S.},
        title = "{Particle acceleration and non-thermal emission in colliding-wind binary systems}",
      journal = {\mnras},
         year = 2021,
        month = jul,
       volume = {504},
       number = {3},
        pages = {4204-4225},
          doi = {10.1093/mnras/stab1107},
archivePrefix = {arXiv},
       eprint = {2104.07399},
 primaryClass = {astro-ph.HE},
       adsurl = {https://ui.adsabs.harvard.edu/abs/2021MNRAS.504.4204P}
}

@INPROCEEDINGS{Bordas2024,
       author = {{Bordas}, P.},
        title = "{Gamma-ray emitting binaries}",
    booktitle = {7th Heidelberg International Symposium on High-Energy Gamma-Ray Astronomy},
         year = 2024,
        month = dec,
        pages = {17},
       adsurl = {https://ui.adsabs.harvard.edu/abs/2024hegr.confE..17B}
}

@INPROCEEDINGS{Chernyakova2020,
       author = {{Chernyakova}, M. and {Malyshev}, D.},
        title = "{Gamma-ray binaries}",
    booktitle = {Multifrequency Behaviour of High Energy Cosmic Sources - XIII. 3-8 June 2019. Palermo},
         year = 2020,
        month = dec,
          eid = {45},
        pages = {45},
          doi = {10.22323/1.362.0045},
archivePrefix = {arXiv},
       eprint = {2006.03615},
 primaryClass = {astro-ph.HE},
       adsurl = {https://ui.adsabs.harvard.edu/abs/2020mbhe.confE..45C}
}

@inproceedings{Dhawan2006,
	adsurl = {http://adsabs.harvard.edu/abs/2006smqw.confE..52D},
	author = {{Dhawan}, V. and {Mioduszewski}, A. and {Rupen}, M.},
	booktitle = {Proceedings of the VI Microquasar Workshop: Microquasars and Beyond},
	publisher = {Como, Italy, PoS(MQW6)052},
	title = {{LS I +61 303 is a Be-Pulsar binary, not a Microquasar}},
	year = 2006}

@article{Moldon2011,
	adsurl = {http://adsabs.harvard.edu/abs/2011ApJ...732L..10M},
	archiveprefix = {arXiv},
	author = {{Mold{\'o}n}, J. and {Johnston}, S. and {Rib{\'o}}, M. and {Paredes}, J.~M. and {Deller}, A.~T.},
	doi = {10.1088/2041-8205/732/1/L10},
	eid = {L10},
	eprint = {1103.1411},
	journal = {\apjl},
	month = may,
	pages = {L10},
	primaryclass = {astro-ph.HE},
	title = {{Discovery of Extended and Variable Radio Structure from the Gamma-ray Binary System PSR B1259-63/LS 2883}},
	volume = 732,
	year = 2011}

@article{Moldon2012,
	adsurl = {http://adsabs.harvard.edu/abs/2012A%26A...548A.103M},
	author = {{Mold{\'o}n}, J. and {Rib\'o}, M. and {Paredes}, J.~M.},
	doi = {10.1051/0004-6361/201219553},
	eprint = {arXiv:astro-ph/1209.6073},
	journal = {\aap},
	month = dec,
	pages = {A103},
	title = {{Periodic morphological changes in the radio structure of the gamma-ray binary LS 5039}},
	volume = 548,
	year = 2012}

@article{massi2001,
	adsurl = {http://adsabs.harvard.edu/abs/2001A%26A...376..217M},
	author = {{Massi}, M. and {Rib{\'o}}, M. and {Paredes}, J.~M. and {Peracaula}, M. and {Estalella}, R.},
	doi = {10.1051/0004-6361:20010953},
	eprint = {astro-ph/0107093},
	journal = {\aap},
	month = sep,
	pages = {217-223},
	title = {{One-sided jet at milliarcsecond scales in LS I +61$^{\circ}$303}},
	volume = 376,
	year = 2001}

@inproceedings{Marcote2014,
	author = {{Marcote}, B. and {Mold{\'o}n}, J. and {Rib\'o}, M. and {Paredes}, J.~M. and {Paragi}, Z.},
	booktitle = {Proceedings of the 12th European VLBI Network Symposium and Users Meeting},
	publisher = {Cagliari, Italy, PoS(EVN 2014)95},
	title = {{The changing morphology of the radio outflow of HESS~J0632+057 along its orbit}},
	year = 2014}

@article{Marcote2015ls5039,
	adsurl = {http://adsabs.harvard.edu/abs/2015arXiv150407253M},
	author = {{Marcote}, B. and {Rib{\'o}}, M. and {Paredes}, J.~M. and {Ishwara-Chandra}, C.~H.},
	eprint = {astro-ph/1504.07253},
	journal = {\mnras},
	month = apr,
	pages = {4578},
	title = {{Physical properties of the gamma-ray binary LS~5039 through low and high frequency radio observations}},
	volume = 451,
	year = 2015}

@ARTICLE{Marcote2016lsi,
       author = {{Marcote}, B. and {Rib{\'o}}, M. and {Paredes}, J.~M. and {Ishwara-Chandra}, C.~H. and {Swinbank}, J.~D. and {Broderick}, J.~W. and {Markoff}, S. and {Fender}, R. and {Wijers}, R.~A.~M.~J. and {Pooley}, G.~G. and et al.},
        title = "{Orbital and superorbital variability of LS I +61 303 at low radio frequencies with GMRT and LOFAR}",
      journal = {\mnras},
         year = 2016,
        month = feb,
       volume = {456},
       number = {2},
        pages = {1791-1802},
          doi = {10.1093/mnras/stv2771},
archivePrefix = {arXiv},
       eprint = {1512.03016},
 primaryClass = {astro-ph.HE},
       adsurl = {https://ui.adsabs.harvard.edu/abs/2016MNRAS.456.1791M}
}

@article{Moldon2011hess,
	adsurl = {http://adsabs.harvard.edu/abs/2011A%26A...533L...7M},
	archiveprefix = {arXiv},
	author = {{Mold{\'o}n}, J. and {Rib{\'o}}, M. and {Paredes}, J.~M.},
	doi = {10.1051/0004-6361/201117764},
	eid = {L7},
	eprint = {1108.0437},
	journal = {\aap},
	month = sep,
	pages = {L7},
	primaryclass = {astro-ph.HE},
	title = {{Revealing the extended radio emission from the gamma-ray binary HESS J0632+057}},
	volume = 533,
	year = 2011}

@ARTICLE{BoschRamon2012,
       author = {{Bosch-Ramon}, V. and {Barkov}, M.~V. and {Khangulyan}, D. and {Perucho}, M.},
        title = "{Simulations of stellar/pulsar-wind interaction along one full orbit}",
      journal = {\aap},
         year = 2012,
        month = aug,
       volume = {544},
          eid = {A59},
        pages = {A59},
          doi = {10.1051/0004-6361/201219251},
archivePrefix = {arXiv},
       eprint = {1203.5528},
 primaryClass = {astro-ph.HE},
       adsurl = {https://ui.adsabs.harvard.edu/abs/2012A&A...544A..59B}
}

@ARTICLE{BoschRamon2011,
       author = {{Bosch-Ramon}, V. and {Barkov}, M.~V.},
        title = "{Large-scale flow dynamics and radiation in pulsar {\ensuremath{\gamma}}-ray binaries}",
      journal = {\aap},
         year = 2011,
        month = nov,
       volume = {535},
          eid = {A20},
        pages = {A20},
          doi = {10.1051/0004-6361/201117235},
archivePrefix = {arXiv},
       eprint = {1105.6236},
 primaryClass = {astro-ph.HE},
       adsurl = {https://ui.adsabs.harvard.edu/abs/2011A&A...535A..20B}
}

@ARTICLE{Seo2018,
       author = {{Seo}, Jeongbhin and {Kang}, Hyesung and {Ryu}, Dongsu},
        title = "{The Contribution of Stellar Winds to Cosmic Ray Production}",
      journal = {Journal of Korean Astronomical Society},
         year = 2018,
        month = apr,
       volume = {51},
       number = {2},
        pages = {37-48},
          doi = {10.5303/JKAS.2018.51.2.37},
archivePrefix = {arXiv},
       eprint = {1804.07486},
 primaryClass = {astro-ph.HE},
       adsurl = {https://ui.adsabs.harvard.edu/abs/2018JKAS...51...37S}
}

@INPROCEEDINGS{Kalyashova2019,
       author = {{Kalyashova}, M.~E. and {Bykov}, A.~M. and {Osipov}, S.~M. and {Ellison}, D.~C. and {Badmaev}, D.~V.},
        title = "{Wolf-Rayet stars in young massive star clusters as potential sources of Galactic cosmic rays}",
    booktitle = {Journal of Physics Conference Series},
         year = 2019,
       series = {Journal of Physics Conference Series},
       volume = {1400},
        month = nov,
    publisher = {IOP},
          eid = {022011},
        pages = {022011},
          doi = {10.1088/1742-6596/1400/2/022011},
archivePrefix = {arXiv},
       eprint = {1910.08602},
 primaryClass = {astro-ph.HE},
       adsurl = {https://ui.adsabs.harvard.edu/abs/2019JPhCS1400b2011K}
}

@ARTICLE{Wang2022,
       author = {{Wang}, Kai and {Zhang}, Hai-Ming and {Liu}, Ruo-Yu and {Wang}, Xiang-Yu},
        title = "{Detection of Diffuse {\ensuremath{\gamma}}-Ray Emission toward a Massive Star-forming Region Hosting Wolf-Rayet Stars}",
      journal = {\apj},
         year = 2022,
        month = aug,
       volume = {935},
       number = {2},
          eid = {129},
        pages = {129},
          doi = {10.3847/1538-4357/ac815e},
archivePrefix = {arXiv},
       eprint = {2207.06583},
 primaryClass = {astro-ph.HE},
       adsurl = {https://ui.adsabs.harvard.edu/abs/2022ApJ...935..129W}
}

@article{Moldon2012lspsr,
	adsurl = {http://adsabs.harvard.edu/abs/2012A%26A...543A..26M},
	archiveprefix = {arXiv},
	author = {{Mold{\'o}n}, J. and {Rib{\'o}}, M. and {Paredes}, J.~M. and {Brisken}, W. and {Dhawan}, V. and {Kramer}, M. and {Lyne}, A.~G. and {Stappers}, B.~W.},
	doi = {10.1051/0004-6361/201219205},
	eid = {A26},
	eprint = {1205.2080},
	journal = {\aap},
	month = jul,
	pages = {A26},
	primaryclass = {astro-ph.HE},
	title = {{On the origin of LS 5039 and PSR J1825-1446}},
	volume = 543,
	year = 2012}

@article{Marcote2018,
	adsurl = {https://ui.adsabs.harvard.edu/\#abs/2018A&A...619A..26M},
	archiveprefix = {arXiv},
	author = {{Marcote}, B. and {Rib{\'o}}, M. and {Paredes}, J.~M. and {Mao}, M.~Y. and {Edwards}, P.~G.},
	doi = {10.1051/0004-6361/201832572},
	eid = {A26},
	eprint = {1809.01119},
	journal = {\aap},
	month = Nov,
	pages = {A26},
	primaryclass = {astro-ph.HE},
	title = {{Refining the origins of the gamma-ray binary 1FGL J1018.6-5856}},
	volume = {619},
	year = 2018}

@ARTICLE{MillerJones2018,
       author = {{Miller-Jones}, J.~C.~A. and {Deller}, A.~T. and {Shannon}, R.~M. and {Dodson}, R. and {Mold{\'o}n}, J. and {Rib{\'o}}, M. and {Dubus}, G. and {Johnston}, S. and {Paredes}, J.~M. and {Ransom}, S.~M. and {Tomsick}, J.~A.},
        title = "{The geometric distance and binary orbit of PSR B1259-63}",
      journal = {\mnras},
         year = 2018,
        month = oct,
       volume = {479},
       number = {4},
        pages = {4849-4860},
          doi = {10.1093/mnras/sty1775},
archivePrefix = {arXiv},
       eprint = {1804.08402},
 primaryClass = {astro-ph.HE},
       adsurl = {https://ui.adsabs.harvard.edu/abs/2018MNRAS.479.4849M}
}

@article{Wu2018,
	adsurl = {https://ui.adsabs.harvard.edu/#abs/2018MNRAS.474.4245W},
	author = {{Wu}, Y.~W. and {Torricelli-Ciamponi}, G. and {Massi}, M. and {Reid}, M.~J. and {Zhang}, B. and {Shao}, L. and {Zheng}, X.~W.},
	doi = {10.1093/mnras/stx3003},
	journal = {\mnras},
	month = Mar,
	pages = {4245-4253},
	title = {{Revisiting LS I +61$^{\circ}$303 with VLBI astrometry}},
	volume = {474},
	year = 2018}

@article{Zabalza2013,
	adsurl = {http://adsabs.harvard.edu/abs/2013A%26A...551A..17Z},
	archiveprefix = {arXiv},
	author = {{Zabalza}, V. and {Bosch-Ramon}, V. and {Aharonian}, F. and {Khangulyan}, D.},
	doi = {10.1051/0004-6361/201220589},
	eid = {A17},
	eprint = {1212.3222},
	journal = {\aap},
	month = mar,
	pages = {A17},
	primaryclass = {astro-ph.HE},
	title = {{Unraveling the high-energy emission components of gamma-ray binaries}},
	volume = 551,
	year = 2013}

@phdthesis{CarreteroCastrillo2025,
	author = {{Carretero-Castrillo}, M.},
	month = oct,
	school = {Universitat de Barcelona},
	title = {{Search and study of massive runaway stars in the Milky Way and impact on high-energy binaries}},
	url = {http://hdl.handle.net/10803/695489},
	year = 2025}

\end{document}